\documentclass{article}
\usepackage{epsfig}
\begin{document}

\title{Determination of the b-mass  using renormalon cancellation}

\author{Carlos Contreras\footnote{ Contribution presented at SILAFAE VI, Lima, Peru, July 12-17 2004 }   \\
 Dept. of Physics, Universidad T\'ecnica Federico Santa Mar\'{\i}a, \\
 Valpara\'{\i}so,Chile\\ USM-TH-160}
\maketitle

\begin{abstract}
Methods of Borel integration to
calculate the binding ground energies and mass
of $b {\bar b}$  quarkonia are presented.
The methods take into account the leading infrared renormalon
structure of the "soft"  binding energy
$E(s)$   and of the   quark pole masses $m_q$, and the  property
that the contributions of  these singularities in
$M(s) = 2 m_q + E(s)$ cancel.   The resummation
formalisms are applied to   quantities which do not
involve renormalon ambiguity, such as ${\overline {\rm MS}}$
mass ${\overline m}_q$ and $\alpha_s(\mu)$.

\end{abstract}

\maketitle

\section{Introduction}
\label{intro}
The calculation
of binding energies, masses of heavy quarkonia $q {\bar q}$ and
another physical parameter
using  renormalon method has attracted the attention recently. The calculations, based on perturbative expansions,
are primarily due to the knowledge of up to ${\rm N}^2{\rm LO}$
term ($\sim\!\alpha_s^3$) of the static quark-antiquark
potential $V(r)$ \cite{Peter:1996ig,Schroder:1998vy}
and partial knowledge of the ${\rm N}^3{\rm LO}$ term there,
and the ultrasoft gluon contributions to a corresponding
effective theory ${\rm N}^3{\rm LO}$ Hamiltonian
\cite{Kniehl:2002br,Brambilla:1999qa,Kniehl:1999ud};
and the knowledge of  the pole mass $m_q$ up to order
$\sim \alpha_s^3$ \cite{Gray:1990yh,Chetyrkin:1999qi}.
Another impetus in these calculations was given by
the observation of the fact that the contributions of the
leading infrared (IR) renormalon singularities (at $b=1/2$)
of the pole mass $m_q$ and of the static potential $V(r)$
cancel in the sum $2 m_q + V(r)$
\cite{Hoang:1998nz,Brambilla:1999xf,Beneke:1998rk}.
Consequently, this cancellation effect must be present also
in the total quarkonium mass $M = 2 m_q + E_{q \bar q}$
\cite{Pineda:2001zq,Lee:2003hh}, or more precisely,
in $M(s) = 2 m_q + E(s)$
where $E(s)$ is the hard$+$soft part of the binding energy,
i.e., the part which includes the contributions
of relative quark-antiquark momenta $|k^0|$, $|{\bf k}| \geq
m_q \alpha_s $, i.e., soft/potential scales (predominant)
and higher hard scales (smaller contributions).
In addition, the binding energy has contribution $E_{q \bar q}(us)$
from the ultrasoft momenta regime
$|k^0|$, $|{\bf k}| \sim m_q \alpha^2_s $. The ultrasoft
contribution is not related to the $b=1/2$ renormalon
singularity, since this singularity has to do with the
behavior of   theory in the region which includes the hard
($\sim m_q$) and soft/potential ($\sim m_q \alpha_s$) scales.

This contribution is based on ref. \cite{CCG:2004} and we present
  numerical calculation of
the binding ground energies $E_{b \bar b}$
(separately the $s$ and the $us$ parts)
and the mass $(2 m_b + E_{b \bar b})$ of the heavy $b {\bar b}$
system, by taking into account the leading IR
renormalon structure of $m_b$ and $E_{b \bar b}(s)$,
and combining some features as:
(a) the mass that we use in the perturbation expansions
is a renormalon-free mass
\cite{Beneke:1998rk,Pineda:2001zq,Bigi:1994em,Hoang:1998ng,Yakovlev:2000pv}
which we choose
to be the ${\overline {\rm MS}}$ mass
${\overline m}_q \equiv {\overline m}_q(\mu = {\overline m}_q)$;
(b) Borel integrations \cite{Lee:2003hh} are used to perform
resummations. Before resummations we perform
separation of the soft/potential ($s$) and ultrasoft ($us$)
part of the binding energies, and apply the renormalon-based
Borel resummation only to the $s$ part. The renormalization
scales used in the Borel resummations are $\mu_h \sim m_q$ (hard scale)
for $2m_{q}$, and $m_{q}\alpha_{s} \leq  {\mu}_{s} < m_{q}$ for $E_{q \bar q}(s)$.
The term corresponding to $E_{q \bar q}(us)$ is evaluated
at $\mu_{us} \sim m_q \alpha^{2}_{s}$ whenever perturbatively possible.
Further, the Borel resummations are performed
in three different ways: (a) using a slightly extended version
of the full bilocal expansion  Refs.~\cite{Lee:2002sn,Lee:2003hh};
(b) using a new ``$\sigma$-regularized'' full
bilocal expansion introduced in the ref. \cite{CCG:2004};
(c) using the form of the Borel transform where the leading IR
renormalon structure is a common factor of the transform
\cite{Caprini:1998wg,Cvetic:2001sn} (we call it $R$-method).
The Borel integrations for both $m_q$ and $E_{q \bar q}(s)$
are performed by the same prescription (generalized principal value PV
\cite{Khuri:bf,Caprini:1998wg,Cvetic:2001sn,Cvetic:2002qf})
so as to ensure the numerical cancellation of the
renormalon contributions in the sum $2 m_q + E_{q \bar q}(s)$.
Furthermore, we demonstrate numerically that in the latter
sum the residues at the renormalons are really
consistent with the renormalon cancellation
when a reasonable factorization scale parameter for the $s$-$us$
separation is used, while they become incosistent with
the aforementioned cancellation when no such separation is used.
The  numerical results allow us to extract the
mass ${\overline m}_b$ from the known $\Upsilon(1S)$
mass of the $b {\bar b}$ system.

\section{Pole mass}
\label{mass}

Here we   calculate   the pole mass $m_q$
in terms of the ${\overline {\rm MS}}$ renormalon-free mass
${\overline m}_q \equiv {\overline m}_q(\mu = {\overline m}_q)$
and of $\alpha_s(\mu,{\overline {\rm MS}})$, using
elements of the renormalization group approach of Ref.~\cite{Pineda:2001zq}
and the bilocal expansion method  Refs.~\cite{Lee:2002sn,Lee:2003hh}.
The ratio $ m_q/{\overline m}_q $ has, divergent, perturbation expansion
in ${\overline {\rm MS}}$ scheme which is at present known
to order $\sim\!\alpha_s^3$
(Ref.~\cite{Gray:1990yh} for $\sim\!\alpha_s^2$;
\cite{Chetyrkin:1999qi} for $\sim\!\alpha_s^3$. Let us define
\label{Sm}
\begin{eqnarray}
S \equiv \frac{m_q}{{\overline m}_q} - 1 & = & \frac{4}{3} a(\mu)
\left[ 1 + a(\mu) r_1(\mu) + a^2(\mu) r_2(\mu) + {\cal O}(a^3) \right]
\ ,
\label{Smexp}
\\
r_1(\mu) & = & \kappa_1 + \beta_0 L_m(\mu)
\ ,
\label{r1}
\\
r_2(\mu) & = & \kappa_2 + ( 2 \kappa_1 \beta_0 + \beta_1) L_m(\mu)
+ \beta_0^2 L_m^2(\mu) \ ,
\label{r2}
\\
(4/3) \kappa_1 & = & 6.248 \beta_0 - 3.739 \ ,
\label{k1}
\\
(4/3) \kappa_2 &= &   23.497 \beta_0^2 + 6.248 \beta_1
+ 1.019 \beta_0 - 29.94 \ ,
\label{k2}
\end{eqnarray}
where $L_m = \ln(\mu^2/{\overline m}_q^2)$, while
$\beta_0 = (11 - 2 n_f/3)/4$ and $\beta_1 = (102 - 38 n_f/3)/16$
are the renormalization scheme independent coefficients
with $n_f = n_{\ell}$ being the number of light active
flavors (quarks with masses lighter than $m_q$).
The natural renormalization scale here is $\mu = \mu_h \sim m_q$
(hard scale).

Let us consider $S$ as   a  function of the running coupling
$a(Q)=\alpha (q)/\pi$, and  the  perturbation
expansion for $S$ reads:
\begin{equation}
S(a(Q))=\sum_{n=0}^{\infty} s_{n}a^{n},
\end{equation}
which has to be summed. Then, one defines the Borel transform as
\begin{equation}
B(b,Q)=\sum_{n=0}^{\infty} s_{n}\frac{b^{n}}{n!}.
\end{equation}
This series has a finite radius of convergence in the $b$-plane,
and we introduce the Borel function $\widetilde{S}(a)$ (Borel Integral) corresponding to
$S(a)$ as
\begin{equation}
\widetilde{S}(a)=\int_{0}^{\infty} db B(b,Q) e^{-b/a}
\end{equation}
 and in our case $B_S(b)$ is known to order $\sim\!b^2$
\begin{equation}
B_{S}(b; \mu) = \frac{4}{3} \left[ 1 + \frac{r_1(\mu)}{1! \beta_0} b +
\frac{r_2}{2! \beta_0^2} b^2 + {\cal O}(b^3) \right] \ .
\label{BSm1}
\end{equation}
It has renormalon singularities at
$b = 1/2, 3/2, 2, \ldots, -1, -2, \ldots$
\cite{Bigi:1994em,Beneke:1994sw,Beneke:1999ui}.
The behavior of $B_S$ near the leading IR renormalon
singularity $b=1/2$ is determined by the resulting
renormalon ambiguity of $m_q$ which has to have the dimensions
of energy and should be renormalization scale and
scheme indepenent -- the only such QCD scale being
$const \times \Lambda_{\rm QCD}$
\cite{Beneke:1994rs}.
The Stevenson scale
$\widetilde \Lambda$ \cite{Stevenson:1981vj}  can be obtained in
terms of the strong coupling parameter
$a(\mu; c_2, c_3, \ldots) = \alpha_s(\mu; c_2, c_3, \ldots)/\pi$,
where $c_j = \beta_j/\beta_0$ ($j \geq 2$) are the
parameters characterizing the renormalization scheme,
by solving the renormalization group equation (RGE)
 \cite{Stevenson:1981vj}
\begin{eqnarray}
\frac{d a(\mu)}{d \ln \mu^2 }&=& - \beta_0 a^2(\mu)(1 + c_1 a(\mu)
+ c_2 a^2(\mu) + \cdots ) \quad \Rightarrow
\label{RGE}
\\
\ln \left( \frac{{\widetilde \Lambda}^2}{ \mu^2 } \right)
&=& \frac{1}{\beta_0} \int_0^{a(\mu)} dx \
\left[ \frac{1}{x^2(1 + c_1 x + c_2 x^2 + \cdots )}-
\frac{1}{x^2(1 + c_1 x)} \right]  \nonumber\\
& &  - \frac{1}{\beta_0 a(\mu)} +
\frac{c_1}{\beta_0} \ln \left( \frac{ 1 + c_1 a(\mu) }{c_1 a(\mu)}
\right) \ \Rightarrow
\label{solRGE}
\\
{\widetilde \Lambda} &=& \mu \
\exp \left( - \frac{1}{2 \beta_0 a(\mu)} \right)
\left( \frac{1 + c_1 a(\mu)}{ c_1 a(\mu)} \right)^{\nu} \nonumber\\
& &
\exp\left[ -  \frac{1}{2 \beta_0} \int_0^{a(\mu)} dx \
\frac{( c_2 + c_3 x + c_4 x^2 + \cdots )}{
( 1 + c_1 x ) ( 1 + c_1 x + c_2 x^2 + \cdots )} \right] \,
\label{tL1}
\end{eqnarray}
where $\nu = c_1/(2 \beta_0) = \beta_1/(2 \beta_0^2)$. Expansion of expression (\ref{tL1}) in powers of
$a(\mu)$ then gives
\begin{equation}
{\widetilde \Lambda} = \mu \exp \left( - \frac{1}{2 \beta_0 a(\mu)}
\right) a(\mu)^{- \nu} c_1^{- \nu} \left[
1 + \sum_{k=1}^{\infty} \ {\widetilde r}_k a^k(\mu) \right] \ ,
\label{tL2}
\end{equation}
where
\begin{eqnarray}
{\widetilde r}_1 &=& \frac{ ( c_1^2 - c_2) }{2 \beta_0} \ ,
\qquad
{\widetilde r}_2 = \frac{1}{8 \beta_0^2} \left[
( c_1^2 - c_2 )^2 - 2 \beta_0 ( c_1^3 - 2 c_1 c_2 + c_3 ) \right] \ ,
\nonumber \\
{\widetilde r}_3 & = & \frac{1}{48 \beta_0^3}  [
( c_1^2 - c_2 )^3 - 6 \beta_0 (c_1^2 - c_2 )(c_1^3 - 2 c_1 c_2 + c_3 )
\nonumber  \\
& & + 8 \beta_0^2 ( c_1^4 - 3 c_1^2 c_2 + c_2^2 + 2 c_1 c_3 - c_4 )] \ .
\end{eqnarray}
The singular part of the Borel transform $B_S(b)$ around $b=1/2$
must have the form
\label{BStc}
\begin{equation}
B_S(b; \mu) =  N_m \pi  \frac{\mu}{ {\overline m}_q }
 \frac{1}{ ( 1 - 2 b)^{1 + \nu} } \left[ 1 +
\sum_{k=1}^{\infty} {\widetilde c}_k ( 1 - 2 b)^k \right]
+ B_{S}^{\rm (an.)}(b; \mu) \ ,
\label{BSrenan}
\end{equation}
\begin{equation}
{\widetilde c}_1 = \frac{ {\widetilde r}_1 }{ (2 \beta_0 ) \nu } \ ,
\quad
{\widetilde c}_2 = \frac{ {\widetilde r}_2}{ (2 \beta_0 )^2 \nu (\nu -1) }
\ , \quad
{\widetilde c}_3 = \frac{ {\widetilde r}_3}
{ (2 \beta_0 )^3 \nu (\nu -1)(\nu - 2) } \ ,
\label{tc}
\end{equation}
and $B_{S}^{\rm (an.)}(b; \mu)$ is analytic on the disk $|b| < 1$.
The ${\overline {\rm MS}}$ coefficients for $ n_f\!=\!4$: $c_1= 1.5400$, $c_2=3.0476$ and $c_3=15.0660$ are
already known \cite{Tarasov:au,vanRitbergen:1997va},
but for $c_4$ we have only
estimates \cite{Ellis:1997sb, Elias:1998bi} obtained
by Pad\'e-related methods. We consider
 $c_4 =40 \pm 60$. Thus, ${\widetilde c}_j$ can be obtained: ${\tilde c_1} = -0.1054 $, ${\tilde c_2} = 0.2736
$ and ${\tilde c_3} = 0.01 \pm 0.17$.

The  bilocal method \cite{Lee:2002sn} consists of taking
in the expansion (\ref{BSrenan}) for the analytic part
$B_{S}^{\rm (an.)}$ a polynomial in powers of $b$, so that
the expansion of $B_S$ around $b\!=\!0$ agrees with
expansion (\ref{BSm1}). For that, the residue parameter
$N_m$ in Eq.~(\ref{BSrenan}) has to be determined.
Using  Refs.~\cite{Cvetic:2003wk}:
\begin{equation}
N_m = \frac{{\overline m}_q}{\mu} \frac{1}{\pi} R_S(b=1/2) \ ,
\label{Nmform}
\end{equation}
and according to (\ref{BSrenan})
\begin{equation}
R_S(b; \mu) \equiv  (1 - 2 b)^{1 + \nu} B_{S}(b; \mu) \ .
\label{RSm}
\end{equation}
Then, the $N_m$  is estimated (see Ref.~\cite{Cvetic:2003wk}),
using $R_S(b)$  TPS
and Pad\'e approximation $[1/1]$:
 \begin{equation}
N_m(n_f\!=\!4) = 0.555 \pm 0.020 \ ,
\label{Nmnf}
 \end{equation}
The bilocal expansion (\ref{BSrenan}) has then for the
analytic part the polynomial
\label{BSan}
\begin{eqnarray}
B_{S}^{\rm (an.)}(b; \mu) & = & h^{(m)}_0 + \frac{h^{(m)}_1}{1! \beta_0} b
+ \frac{h^{(m)}_2}{2! \beta_0^2} b^2 \ ,
\label{BSanexp}
\\
h^{(m)}_k & = & \frac{4}{3} r_k  - \pi N_m \frac{\mu}{{\overline m}_q}
(2 \beta_0)^k \sum_{n=0}^3 {\widetilde c}_n
\frac{ \Gamma ( \nu + k + 1 - n) }{ \Gamma(\nu + 1 - n) } \ ,
\label{hms}
\end{eqnarray}
where, by convention, $r_0 = {\widetilde c}_0 = 1$.
Then, the bilocal formula, is
\begin{equation}
B_S(b; \mu)^{\rm (biloc.)}   =   N_m \pi  \frac{\mu}{ {\overline m}_q }
 \frac{1}{ ( 1 - 2 b)^{1 + \nu} } \left[ 1 +
\sum_{k=1}^3{\widetilde c}_k ( 1 - 2 b)^k \right]
+ \sum_{k=0}^2 \ \frac{h_k^{(m)}}{k! \ \beta_0^k} \ b^k \ .
\label{BSbiloc}
\end{equation}
Applying the (generalized) principal value (PV)
prescription for the Borel integration
\begin{equation}
S(a) = \frac{1}{\beta}_0 {\rm Re}
\int_{\pm i \varepsilon}^{\infty \pm i \varepsilon} \ db
\ \exp \left( - \frac{b}{\beta_0 a(\mu)} \right) \ B_S(b; \mu) \ ,
\label{BSint}
\end{equation}
we obtain the pole mass $m_q$ in terms of the mass ${\overline m}_q$.
The numerical integration is performed, using the
Cauchy theorem ( Refs.~\cite{Cvetic:2001sn}).

\begin{figure}[htb]
\begin{center}
\begin{minipage}[b]{.49\linewidth}
 \centering\epsfig{file=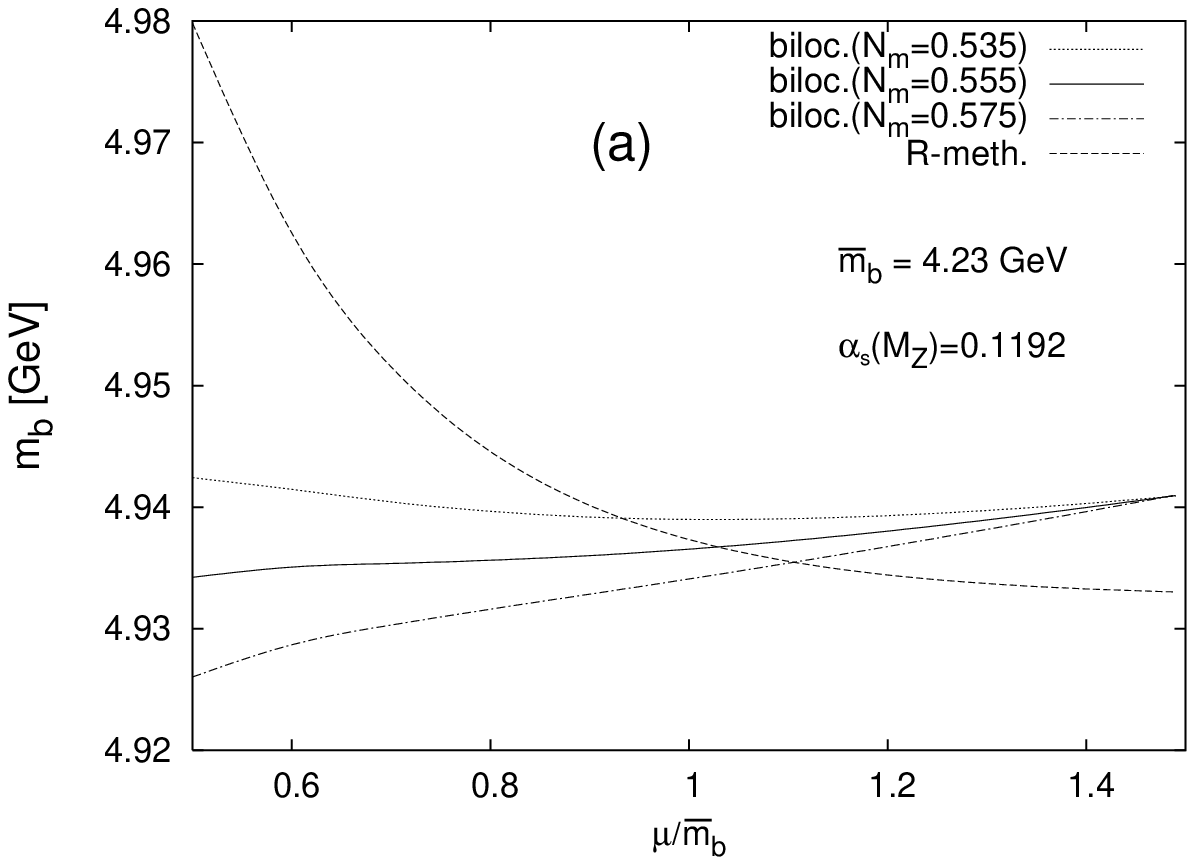,width=\linewidth}
\end{minipage}
\end{center}
 \vspace{0.2cm}
\caption{\footnotesize The (PV) pole mass
of the   bottom   quark, as function of the
renormalization scale $\mu$. The input parameters
used were ${\overline m}_b = 4.23$ GeV and  the residue parameter values Eq.~(\ref{Nmnf}). The reference value for $\alpha_s$
(in ${\overline {\rm MS}}$) was taken to be $\alpha_s(m_{\tau}) = 0.3254$
(\cite{Cvetic:2001ws}) corresponding to $\alpha_s(M_Z) = 0.1192$.}
\label{mqmu.fig}
\end{figure}

In Figs.~\ref{mqmu.fig} (a) we present the resulting (PV) pole
masses of the $b$, as function of the renormalization
scale $\mu$. The spurious $\mu$-dependence is very weak.
In addition, results of another method (``R''-method )
are presented in Figs.~\ref{mqmu.fig} (a),  with the $\mu$-dependence
stronger in the low-$\mu$ region ($\mu/{\overline m}_q < 1$).
The R-method (Refs.~\cite{Caprini:1998wg,Cvetic:2001sn}) consists in
the Borel integration of the function (\ref{RSm})
\begin{equation}
S = \frac{1}{\beta_0} {\rm Re}
\int_{\pm i \varepsilon}^{\infty \pm i \varepsilon} \ db
\ \exp \left( - \frac{b}{\beta_0 a(\mu)} \right)
\ \frac{R_S(b; \mu)}{(1 - 2 b)^{1 + \nu}} \ ,
\label{BRSint}
\end{equation}
where for $R_S(b)$ the corresponding (NNLO) TPS is used.
When we take ${\overline m}_b = 4.23$ GeV  and we vary the values of the residue parameter $N_m$, the bilocal method gives,
at $\mu/{\overline  m}_q = 1$, variation
$\delta m_b = \mp 3$ MeV.
When the central values of $N_m$  are used, the
variation of the obtained values of $m_q$ with $\mu$,
when $\mu/{\overline m}_q$ grows from 1.0 to 1.5, is
about $5$ MeV  for $m_b$
(for $R$-method: $4$ MeV). When $c_4$ is varied, the variation is about $\mp\!2$ and $\mp\!1$ MeV
for $m_b$. The uncertainty in $\alpha_s$
can be taken as
$\alpha_s(m_{\tau}) = 0.3254 \pm 0.0125$ \cite{Cvetic:2001ws},
corresponding to $\alpha_s(M_Z) = 0.1192 \pm 0.0015$.
This uncertainty is by far the major source in the variation
of the pole masses: ($\delta m_b)_{\alpha_s} = ^{+135}_{-148}$ MeV
for bilocal method ($^{+137}_{-150}$ MeV for $R$-method).

The natural renormalization scale $\mu$ here is
a hard scale $\mu\!\sim\!{\overline m}_q$, and will be
denoted later in this work as $\mu_m$ in order to distinguish
if from the ``soft'' renormalization scale $\mu$ used in the
analogous renormalon-based resummations of the
(hard$+$)soft binding energy $E_{q \bar q}(s)$
(${\overline m}_{q} > \mu \geq {\overline m}_{q} \alpha_s$)
in Sec.~\ref{Eqq}. The fact that the two renormalization scales
are different does not affect the mechanism of the
($b\!=\!1/2$) renormalon cancellation in the bilocal calculations of
the meson mass $(2 m_q\!+\!E_{q \bar q}(s))$, because the
renormalon ambiguity in each of the two terms is
renormalization scale independent $\sim\!{\widetilde \Lambda}$,
as seen by Eqs.~(\ref{tL2})--(\ref{BStc}).
On the other hand, if $R$-type methods (\ref{BRSint})
[cf.~also Eq.~(\ref{Nmform})] are applied
for the resummations of $2 m_q$ and $E_{q \bar q}(s)$, the
renormalon ambiguities are renormalization scale independent
in the approximation of the one-loop RGE running, and the
renormalon cancellation is true at this one--loop level.

\section{Separation of the soft and ultrasoft contributions}
\label{sep}

The perturbation expansion of the (hard $+$ soft $+$ ultrasoft)
binding energy $E_{q \bar q}$ of the $q {\bar q}$ heavy quarkonium
vector ($S=1$) or scalar ($S=0$) ground state ($n=1, \ell=0$)
up to the ${\rm N}^3 {\rm LO}$ ${\cal O}(m_q \alpha_s^5)$
was given in \cite{Penin:2002zv}
The reference mass scale used was the pole mass $m_q$.
The ground state energy expansion has the form
\begin{eqnarray}
E_{q \bar q} & = & - \frac{4}{9} m_q \pi^2 a^2(\mu)
{\Big \{} 1 + a(\mu) \left[ k_{1,0} + k_{1,1} L_p(\mu) \right] \nonumber\\
&& +
a^2(\mu) \left[ k_{2,0} + k_{2,1} L_p(\mu) + k_{2,2} L_p^2(\mu) \right]
\nonumber\\
&& + a^3(\mu) \left[ k_{3,0} + k_{3,1} L_p(\mu) + k_{3,2} L_p^2(\mu)
+ k_{3,3} L_p^3(\mu) \right] + {\cal O}(a^4) {\Big \}} \,
\label{Eqqexp}
\end{eqnarray}
where
\begin{eqnarray}
L_p(\mu) &=& \ln \left( \frac{\mu}{\frac{4}{3} m_q \pi a(\mu) } \right) \ .
\label{Lp}
\end{eqnarray}
The   coefficients
$k_{i,j}$ of perturbation expansion (\ref{Eqqexp})   of the quarkonium ($n=1$; $\ell=0$; $S=1$ or $0$)
are given below. \cite{Titard:1993nn,Czarnecki:1997vz,Melnikov:1998ug,Penin:1998zh}, ~\cite{Penin:2002zv} --
For  $N_c=3$:
\label{kij}
\begin{eqnarray}
k_{1,1} & = & 4 \beta_0 \ ,
\qquad
k_{1,0} =  \left( \frac{97}{6} - \frac{11}{9} n_f \right) \ ,
\label{k1s}
\\
k_{2,2} & = & 12 \beta_0^2 \ ,
\qquad
k_{2,1} = \frac{927}{4} - \frac{193}{6} n_f + n_f^2  \ ,
\label{k2221}
\\
k_{2,0} & = & 361.342 - 40.9649 \ n_f + 1.16286 \ n_f^2  - 11.6973 \ S (S+1) \ ,
\label{k20}
\\
k_{3,3} & = & 32 \beta_0^3 \ ,
\qquad
k_{3,2} = \frac{4521}{2} - \frac{10955}{24} n_f + \frac{1027}{36} n_f^2 -
\frac{5}{9} n_f^3 \ ,
\label{k3332}
\\
k_{3,1} & = & 7242.3 - 1243.95 \ n_f + 69.1066 \ n_f^2  - 1.21714 \ n_f^3 \nonumber\\
&&
+ \frac{\pi^2}{2592} ( - 67584 + 4096 \ n_f ) \ S (S+1) \ ,
\label{k31}
\\
k_{3,0} & = & {\bigg [} \left( 7839.82 - 1223.68 \ n_f + 69.4508 \ n_f^2
- 1.21475 \ n_f^3 \right)
\nonumber \\
&& + (- 109.05 + 4.06858 \ n_f) \ S(S+1) \nonumber \\
&&
- \frac{\pi^2}{18}
\left( -1089 + 112 \ S(S+1) \right) \ln \left( a(\mu) \right)
+ 2 \frac{a_3}{4^3}
{\bigg ]} \ ,
\label{k30}
\end{eqnarray}
Here, $a_3$  have been estimated  in  Ref.~\cite{Cvetic:2003wk},
obtained from the condition of renormalon cancellation
in the sum $(2 m_q + V_{q \bar q}(r))$
\label{a3nf}
\begin{equation}
\frac{1}{4^3} \ a_{3}(n_f\!=\!4)   \approx
86. \pm 23. \ .
\end{equation}
The coefficients $k_{i,j}$ in the expansion (\ref{Eqqexp})
originate from quantum effects from various scale regimes
of the participating particles:
(a) the hard scales ($\sim\!m_q$);
(b) the soft and potential scales
where the three momenta are $|{\bf q}|\!\sim\!m_q \alpha_s$
($|q^0|\!\sim\!m_q \alpha_s$ in the soft
and $|q^0|\!\sim\!m_q \alpha_s^2$ in the potential regime);
(c) ultrasoft scales where
$|q^0|$ and $|{\bf q}|$ are both $\sim\!m_q \alpha_s^2$.
The coefficients are dominated by the soft scales;
the hard scales start contributing at the NNLO \cite{Kniehl:2002br}
and are numerically smaller. For this reason, we will usually refer to the combined soft and hard
regime contributions to the binding energy as simply
soft ($s$) contribution $E_{q\bar q}(s)$.
Strictly speaking, it is only the pure soft regime that
contributes to the $b=1/2$ renormalon.
However, for simplicity,
we will resum the hard$+$soft contributions $E_{q \bar q}(s)$
together. This will pose no problem,
since the hard regime, being clearly perturbative,
is not expected to deteriorate the convergence properties
of the series for $E_{q \bar q}(s)$. The natural
renormalization scale $\mu$ in the resummations of $E_{q \bar q}(s)$
is expected to be closer to the soft scale
($m_q \leq  \alpha_s  \mu < m_q$).

On the other hand, the ${\rm N}^3 {\rm LO}$ coefficient
$k_{3,0}$ obtains additional contributions from the
from the ultrasoft ($us$) regime.
The leading ultrasoft contribution comes from the exchange of
an ultrasoft gluon in the heavy quarkonium
\cite{Kniehl:1999ud,Brambilla:1999xf}. It consists of two parts:
\begin{enumerate}
\item
The retarded part, which cannot be interpreted in terms of
an instantaneous interaction
\begin{equation}
\frac{1}{\pi^3} k_{3,0}(us,{\rm ret.}) =
- \frac{2}{3 \pi} \left( \frac{4}{3} \right)^2 L_1^E \approx + 41.014 \ ,
\label{k30usr}
\end{equation}
where $L_1^E \approx - 81.538$ is the QCD Bethe logarithm
-  see Refs.~\cite{Kniehl:1999ud,Kniehl:2002br}.
\item
The non-retarded part can be calculated as expectation value
of the $us$ effective Hamiltonian ${\cal H}^{us}$ in the
Coulomb (i.e., leading order) ground state $|1 \rangle$,
where ${\cal H}^{us}$ (in momentum space) was derived in
Refs.~\cite{Kniehl:1999ud,Kniehl:2002br}. Direct calculation of the
expectation value, here in coordinate space, then gives:
\label{k30usnr}
\begin{eqnarray}
\frac{1}{\pi^3} k_{3,0}(us,{\rm nonret.})&=&
- \frac{9}{4 \pi^5}  \frac{1}{m_q a^5(\mu)}
\langle 1 | {\cal H}^{us} | 1 \rangle =
\frac{2}{ \pi^5 m_q a^4(\mu)}  [ \nonumber \\
& &
\frac{1}{2} \ln \frac{\mu_f^2}{(E_1^C)^2}
  + \frac{5}{6} - \ln 2  ]}
 \times {\Big \{
- \frac{27 \pi^3}{8} a^3(\mu) \langle 1 | \frac{1}{r} | 1 \rangle  \nonumber \\
& &
- 17 \pi^2 \frac{a^2(\mu)}{m_q} \langle 1 | \frac{1}{r^2} | 1 \rangle
 + \frac{4 \pi^2}{3} \frac{a(\mu)}{m_q^2}
\langle 1 | \delta({\bf r}) | 1 \rangle  \nonumber \\
& & +
3 \pi \frac{a(\mu)}{m_q^2} \langle 1 | \{ \Delta_{\bf r}, \frac{1}{r}
\} | 1 \rangle {\Big \}} \ ,
\label{k30usnr1}
\\
 & =&  - 14.196 \left[ \ln \left( \frac{ \mu_f}{m_q \alpha_s^2(\mu)}
\right) + 0.9511 \right] \ .
\label{k30usnr2}
\end{eqnarray}
Here, $E_1^C = -(4/9) m_q \alpha_s^2(\mu)$ is the Coulomb
energy of the state $|1 \rangle$, and $\mu_f$ is the factorization
energy between the soft ($\!\sim m_q \alpha_s$) and ultrasoft
($\!\sim m_q \alpha^2_s$) scale.
\end{enumerate}

The $s$--$us$ factorization scale $\mu_f$ can be
estimated as being roughly in the middle between
the $s$ and $us$ energies on the logarithmic scale
\cite{Cvetic:2003wk}
\begin{equation}
\mu_f \left[ \approx (E_{\rm S} E_{\rm US})^{1/2} \right] = \kappa \ m_q
\alpha_s(\mu_s)^{3/2} \ ,
\label{muf}
\end{equation}
where $\kappa \sim 1$ and $\mu_s \approx E_{\rm S}$ ($\!\mu$).
Therefore, the ultrasoft part of the ${\rm N}^3 {\rm LO}$
coefficient $k_{3,0}$ can be rewritten, by Eqs.~(\ref{k30usr}),
(\ref{k30usnr}) and (\ref{muf}), in terms of the
$s$--$us$ parameter $\kappa$ as
\begin{equation}
\frac{1}{\pi^3} k_{3,0}(us) = 27.512 + 7.098 \ln (\alpha_s(\mu_s))
- 14.196 \ln ( \kappa ) \ .
\label{k30us}
\end{equation}
The soft scale $\mu_s$ appearing here will be fixed
by the condition $\mu_s = (4/3) {\overline m}_q \alpha_s(\mu_s)$.

The formal perturbation expansions for the separate
soft and ultrasoft parts of the ground state binding energy (\ref{Eqqexp})
are then
\label{Eqqsus}
\begin{eqnarray}
E_{q \bar q}(s) & = & - \frac{4}{9} m_q \pi^2 a^2(\mu)
{\Big \{} 1 + \sum_{i=1}^2 a^i(\mu) \sum_{j=0}^i k_{i,j}
L_p(\mu)^j\nonumber \\
& &
+ a^3(\mu) \sum_{j=1}^3 k_{3,j} +
a^3(\mu)  \left[ k_{3,0}\!-\!k_{3,0}(us) \right] + {\cal O}(a^4)
{\Big \}} \ ,
\label{Eqqs}
\\
E_{q \bar q}(us) & = & - \frac{4}{9} m_q \pi^2 a^2(\mu)
{\big \{}  a^3(\mu) k_{3,0}(us)  + {\cal O}(a^4)
{\big \}} \ .
\label{Eqqus}
\end{eqnarray}
The energy $E_{q \bar q}(s)$ (\ref{Eqqs})
contains the leading IR renormalon effects,
and $E_{q \bar q}(us)$ (\ref{Eqqus}) does not.
In these expressions, the common factor is the soft scale
$\mu_p(\mu) = (4/3) m_q \alpha_s(\mu)$ which is also present
as the reference scale in the logarithms
$L_p(\mu) = \ln (\mu/\mu_p(\mu) )$ appearing with the coefficients
$k_{i,j}$ (when $j \geq 1$) in Eqs.~(\ref{Eqqexp}), (\ref{Lp}).

We will re-express $m_q$ everywhere in $E_{q \bar q}$ with the renormalon-free
mass ${\overline m}_q$, and will
consider the dimensionless soft-energy quantity
$E_{q \bar q}(s)/{\overline m}_q$.

Thus, we will divide the soft binding energy
with the quantity ${\overline \mu}({\widetilde \mu})
= (4/3) {\overline m}_q \alpha_s({\widetilde \mu})$, where
${\widetilde \mu}$ can be any soft scale.
We will fix this scale by the condition
${\widetilde \mu} = (4/3) {\overline m}_q  \alpha_s({\widetilde \mu})$
($\Rightarrow {\widetilde \mu}  = \mu_s$).
Further, in the logarithms $L_p(\mu)$ we
express the pole mass $m_q$ in terms of ${\overline m}_q$
and powers of $a(\mu)$ (cf.~Sec.~\ref{mass}),
and the powers of logarithms $\ln^k[a(\mu)]$ we
re-express in terms of $\ln^k [a(\widetilde \mu)]$.
This then results in the following soft binding energy
quantity $F(s)$ to be resummed
\begin{eqnarray}
F(s) & \equiv &  - \frac{9}{4 \pi^2}
\frac{E_{q \bar q}(s)}{ {\overline m}_q a(\widetilde \mu) }
= a(\mu) \left[ 1 + a(\mu) f_1 + a^2(\mu) f_2 + a^3(\mu) f_3
+ {\cal O}(a^4) \right] \ ,
\label{Fs}
\end{eqnarray}
where the coefficients $f_j$ depend on $\ln a(\widetilde \mu)$
and on three scales: the renormalization
scale $\mu$ ($\geq m_q \alpha_s$),
the (fixed) soft scale $\widetilde \mu$,
and ${\overline m}_q$. The coefficient $f_3$ depends,
in addition, on the parameters $\kappa$ (\ref{muf})-(\ref{k30us}),
$\mu_s$, and $a_3$ (\ref{a3nf}).
The coefficients $f_j$ are written explicitly in Appendix \ref{app:sbe}.
The $b=1/2$ renormalon in the quantity $F(s)$ is then of the type
of the renormalon of the pole mass $m_q$ discussed in
the previous Sec.~\ref{mass}.

However, if we divided in Eq.~(\ref{Fs}) by $m_q$ instead of
${\overline m_q}$ and at the same time used in the
resulting $f_j$-coefficients $\ln m_q$, the numerical
resummations of $F(s)$ by methods of Sec.~\ref{Eqq} would
give us values for $E_{q \bar q}(s)$ different usually by
not more than ${\cal O}(10^1 {\rm MeV})$ (we checked this
numerically). We will briefly refer to these approaches later in
this Section as ``pole mass'' approaches. A version of such
pole mass bilocal approach was applied in Ref.~\cite{Lee:2003hh}
for resummation of the unseparated $E_{q \bar q}(s\!+\!us)$.

The ultrasoft part (\ref{Eqqus}), on the other hand,
has no $b = 1/2$ renormalon. The mass scale used there
should also be renormalon free (${\overline m}_q$).
The renormalization scale $\mu$ there should be adjusted downward
to the typical $us$ scale of the associated process
$\mu \mapsto \mu_{us}$ ($\sim\! m_q \alpha_s^2$)
in order to come closer to a realistic estimate
\begin{eqnarray}
E_{q \bar q}(us) & \approx & - \frac{4}{9} {\overline m}_q \pi^2 k_{3,0}(us) a^5(\mu_{us}) \ .
\label{Eqqus2}
\end{eqnarray}

\section{
Mass ${\overline m}_b$ from the known mass of the vector $b {\bar b}$
[$\Upsilon (1S)$]}
\label{Eqq}

The soft binding energy
quantity to be resummed  is $F(s)$ of Eq.~(\ref{Fs}).
However, in the ${\rm N}^3{\rm LO}$ coefficient $f_3$ we have
dependence on   $a_3$, and on the $s$-$us$ factorization scale
parameter $\kappa\!\sim\!1$. Then, the value of  $\kappa$ (\ref{muf})
is obtained requiring that the residue parameter values
be reproduced from the Borel transform of the soft binding
energy quantity $F(s)$ of Eq.~(\ref{Fs}).

Similarly as in Eq.~(\ref{BSrenan}), we have
\begin{equation}
B_{F(s)}(b; \mu) = N_m \frac{9}{2 \pi}
\frac{\mu}{ {\overline m}_q a(\widetilde \mu) }
\frac{1}{ ( 1\!-\!2 b)^{1 + \nu} } \left[ 1 +
\sum_{k=1}^{\infty} {\widetilde c}_k ( 1\!-\!2 b)^k \right]
+ B_{F(s)}^{\rm (an.)}(b; \mu)\,
\label{BFrenan}
\end{equation}
where the factor in front of the singular part
was determined by the condition of renormalon cancellation
of the sum $2 m_q + E_{q \bar q}(s)$. We now define in analogy
with Eq.~(\ref{RSm})
\begin{equation}
R_{F(s)}(b; \mu; \mu_f) = (1 - 2 b)^{1 + \nu} B_{F(s)}(b;\mu; \mu_f) \ .
\label{RFs}
\end{equation}
Here we denoted,  explicitly the dependence
on the factorization scale $\mu_f$ and
\begin{equation}
N_m = \frac{2 \pi}{9} \frac{ {\overline m}_q a(\widetilde \mu) }{\mu}
R_{F(s)}(b;\mu;\mu_f) {\big |}_{b=1/2} \ .
\label{NmEqq}
\end{equation}
Theoretically, $R_{F(s)}(b)$ should be a function
with only a weak singularity (cut) at $b=1/2$, and the
nearest pole at $b=3/2$ ( \cite{Aglietti:1995tg}).
Resummations such as Pad\'e approximations (PA's)
are applied, then
  $R_{F(s)}[2/1](b)$ has physically acceptable
pole structure $|b_{\rm pole}| \geq 1$ for most of the
values of $\mu \geq m_q \alpha_s$ and $\kappa \sim 1$.
and the residue parameter $N_m$ is
reasonably stable under the variation of $\mu$.

\begin{figure}[htb]
\begin{minipage}[b]{.49\linewidth}
 \centering\epsfig{file=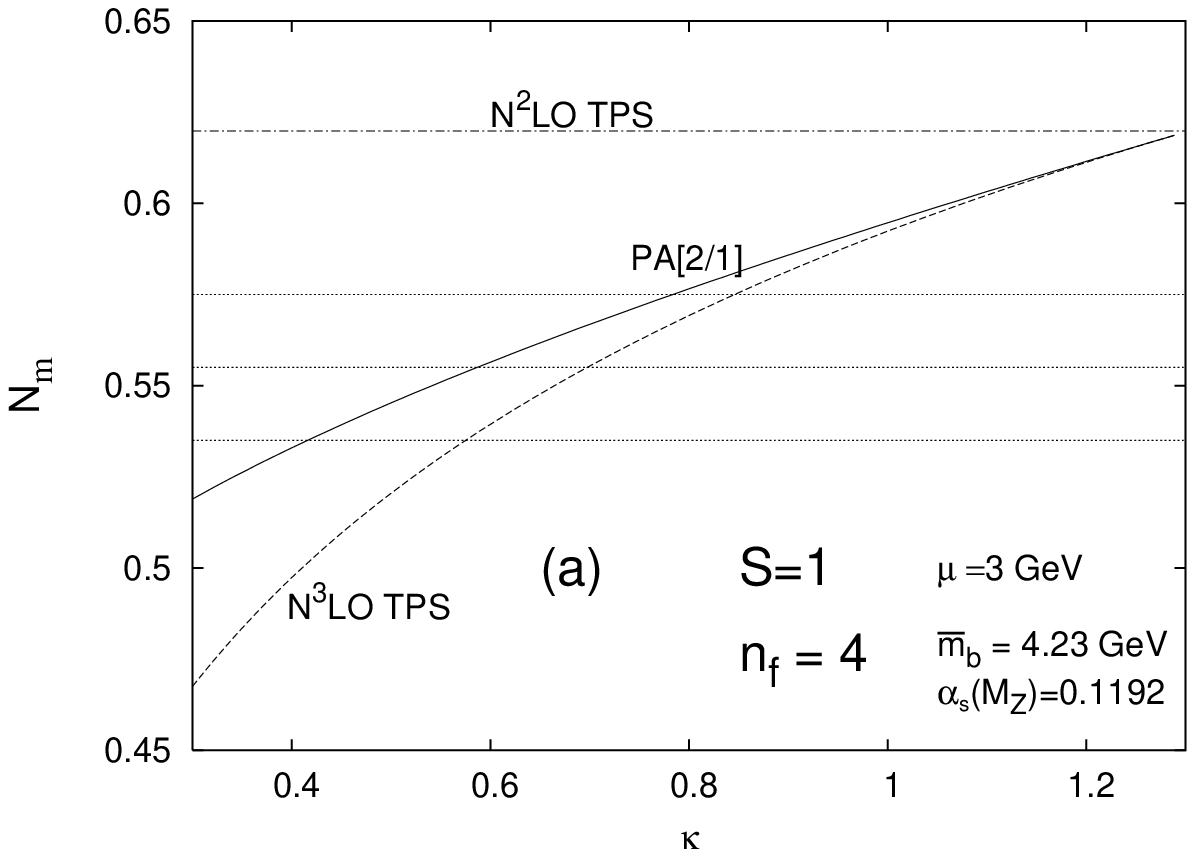,width=\linewidth}
\end{minipage}
\begin{minipage}[b]{.49\linewidth}
 \centering\epsfig{file=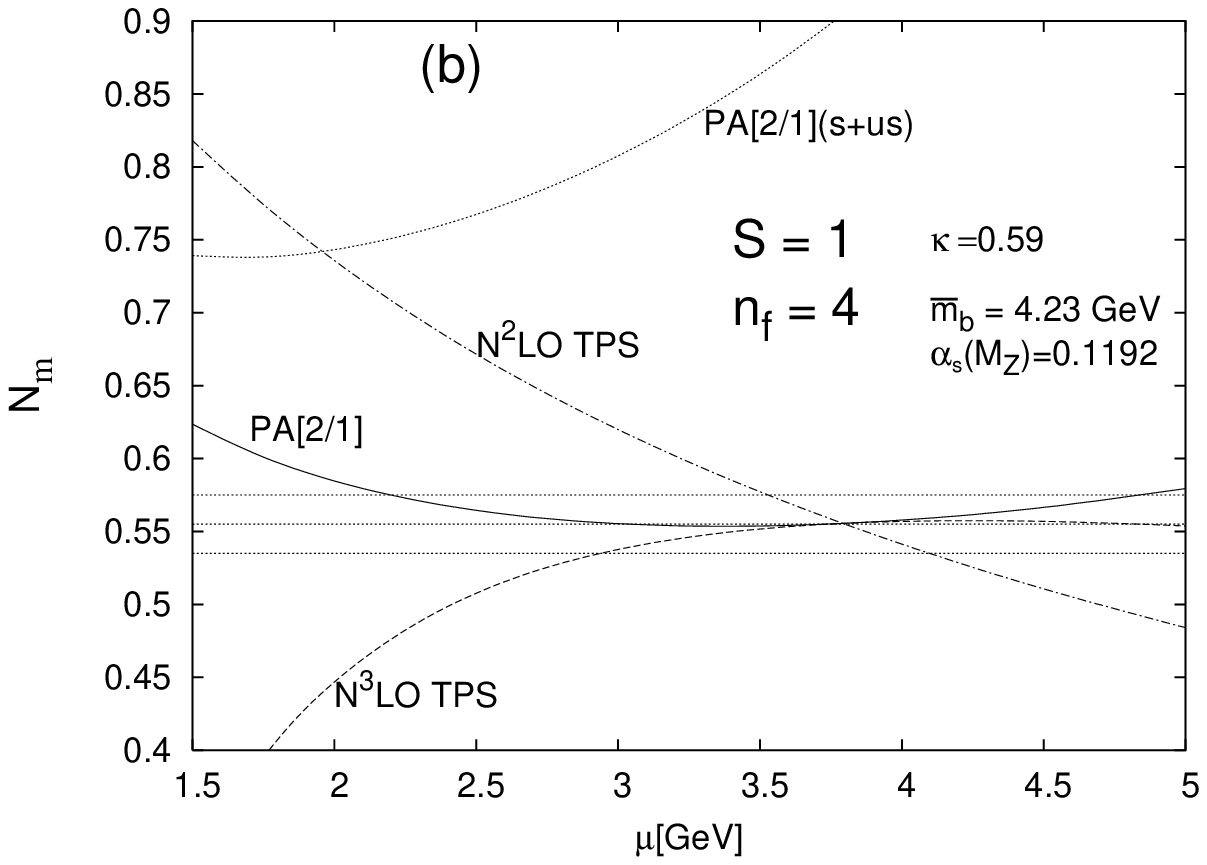,width=\linewidth}
\end{minipage}
\vspace{0.2cm}
\caption{The residue parameter value $N_m$ as calculated from the
soft part of the binding energy of the bottonium according to
Eq.~(\ref{NmEqq}), (a) as a function of the $s$-$us$ factorization
scale parameter $\kappa$ (\ref{muf}), at $\mu=3$ GeV;
(b) as a function of the renormalization scale $\mu$, at
$\kappa = 0.59$. Further explanations given in the text.
In Fig.~(a), the known values  of $N_m$ are denoted as dotted
horizontal lines.}
\label{Nmbb.fig}
\end{figure}

In Fig.~\ref{Nmbb.fig}(a) we show the
dependence of $N_m$ on $\kappa$, at a typical
$\mu$ value $\mu\!=\!3$ GeV, for the $b \bar b$ system.
The  know value  of $N_m$ is
obtained by the $R_{F(s)}[2/1](b\!=\!1/2)$ expression at
$\kappa \approx 0.59$. In Fig.~\ref{Nmbb.fig}(b) we
present, for $\kappa = 0.59$, the dependence of
calculated $N_m$ on the renormalization scale $\mu$.
There, we include also the ($[2/1]$-resummed) curve for
the case when no separation of the $s$ and $us$ parts
of the energy is performed. In that case, the
obtained values of $N_m$ are unacceptable.
The other values of the input parameters are chosen
to have the $b \bar b$ ``central'' values:
$a_3/4^3 = 86$; ${\overline m}_b = 4.23$ GeV;
${\widetilde \mu} = 1.825$ GeV ($\approx \mu_s$)
and $\alpha_s(\widetilde \mu; n_f\!=\!4) = 0.3263 (\approx
\alpha_s(\mu_s; n_f\!=\!4) = 0.326$)
[from: $\alpha_s(m_{\tau}; n_f\!=\!3) = 0.3254$, i.e.,
$\alpha_s(M_Z) = 0.1192$ \cite{Cvetic:2001ws}.
For the RGE running, we   use four-loop ${\overline {\rm MS}}$
$\beta$-function (TPS).

Variation $N_m = 0.555 \pm 0.020$ [$b \bar b$]
implies $\kappa = 0.59 \pm 0.19$.
If, on the other hand, $a_3$ parameter is varied,
 then for $b \bar b$ $\kappa = 0.59 \mp 0.06$.
Thus, the value of $s$-$us$ factorization scale parameter $\kappa$
is influenced largely by the allowed values
of the renormalon residue parameter, and significantly less
by the allowed values $a_3$
of the ${\rm N}^3{\rm LO}$ coefficient of the static $q \bar q$
potential. Therefore, we will consider the variations
of $N_m$  and of $\kappa$ to be
related by a one-to-one relation, while the variations
of $a_3$  will be considered as independent.

In this way, we have the following value  for the $s$-$us$
factorization scale parameter:
\begin{equation}
N_m = 0.555 \pm 0.020 \ \Rightarrow \
\kappa   =   0.59 \pm 0.19
\quad (n_f = 4, \ S = 1) \
\end{equation}
and thus we obtain the ${\rm N}^3{\rm LO}$ TPS (\ref{Fs})
for the soft part of the ground binding energy.
Now with  the value of $\kappa$,
we can perform the resummation of the soft part of the
ground binding energy.
The full bilocal method \cite{Lee:2002sn,Lee:2003hh}
can be performed as in Sec.~\ref{mass},
Eqs.~(\ref{BSbiloc}) and (\ref{BSint}).
Therefore
\begin{equation}
B_{F(s)}^{(\rm biloc.)}(b;\mu)   =
N_m \frac{9}{2 \pi}
\frac{\mu}{ {\overline m}_q a(\widetilde \mu) }
\frac{1}{ ( 1\!-\!2 b)^{1 + \nu} } \sum_{k=0}^3 {\widetilde c}_k (1 - 2 b)^k
+ \sum_{k=0}^3 \ \frac{h_k}{k! \ \beta_0^k} \ b^k \ ,
\label{Fsbiloc}
\end{equation}
where the coefficients
$h_k$ in the expansion of the analytic part are     known
up to order $k=3$
\begin{equation}
h_k = f_k - N_m \ \frac{9}{2 \pi} \
\frac{\mu}{ {\overline m}_q a({\widetilde \mu}) } \
(2 \beta_0)^k \sum_{n=0}^3 \
{\widetilde c}_n
\frac{ \Gamma ( \nu + k + 1 - n) }{ \Gamma(\nu + 1 - n) }
\quad (k=0,1,2,3) \ .
\label{hss}
\end{equation}
  The result
  have   spurious $\mu$-dependence, and  for the of the Borel transformation
  turns out to have a problematic behavior:   the obtained pole of  $B_{F(s)}^{\rm (an.)}[2/1](b)$ is
  unacceptably small in size:
$|b_{\rm pole}| \leq 1/2$. Theoretically, $B_{F(s)}^{\rm (an.)}(b)$
should have the nearest pole at $b=3/2$ \cite{Aglietti:1995tg}.
The reason for this problem
appears to lie in the specific truncated form of the
singular part taken in the bilocal method.
While the latter part describes well the behavior of the
transform near $b=1/2$, it influences apparently strongly
the coefficients $h_k$ and thus the analytic part,
so that no reliable resummation of that part (apart from
TPS) can be done.   This problem can be alleviated by
introducing   a ``form'' factor
which suppresses that part away from $b \approx 1/2$,
but keeps it unchanged at $b \approx 1/2$. We choose
  a Gaussian type of function, and the ``$\sigma$-regularized''
bilocal expressions for the Borel transform is
\begin{eqnarray}
B_{F(s)}^{(\sigma)}(b;\mu) & = & N_{m}   \frac{9}{2 \pi} \frac{\mu}{ {\overline m}_{q} a( \widetilde \mu )} \
\frac{1}{ ( 1\!-\!2 b) ^{1 + \nu } }  [1 + {\widetilde c}_{1} (1\!-\!2b)  \nonumber  \\
&  & +   ( {\widetilde c}_{2} + \frac{1}{8 \sigma^{2}}   )
(1\!-\!2b)^{2}
+  ( {\widetilde c}_{3} + \frac{{\widetilde c}_{1}}{8 \sigma^{2}}  )(1\!-\!2b)^{3}   ]  \nonumber  \\
&  & \times \exp \left[ - \frac{1}{8 \sigma^{2}} (1\!-\!2 b)^2 \right]
+
\sum_{k=0}^{3} \ \frac{1}{k! \ \beta_{0}^{k}} \ h^{(\sigma)}_{k}  b^{k}  \ .
\label{Fssig}
\end{eqnarray}
The corrective terms $1/(8 \sigma^2)$ and
$ {\widetilde c}_1/(8 \sigma^2)$ in the
coefficients of Eq.~(\ref{Fssig}) appear to ensure the correct known behavior
of the renormalon part up to order $\sim\!(1 - 2b)^{-\nu + 2}$.
Numerical analysis indicate that $\sigma$ is between zero and one.
Namely, when $\sigma$ decreases from $\sigma\!=\!\infty$ to about
$\sigma \approx 0.3$-$0.4$, the
value of the pole of the $[2/1]$ Pad\'e-resummed analytic part
$B_{F(s)}^{({\rm an.} \sigma)}(b)$ of Eq.~(\ref{Fssig})
gradually turns acceptable ($|b_{\rm pole}| > 1$) and
rather stable when the renormalization
scale $\mu$ varies in the interval $[m_q \alpha_s, m_q]$
(except close to $\mu \approx m_q \alpha_s$). When the value of $\sigma$
falls below $0.3$, the analytic part starts showing
erratic behavior again and the Borel resummation
gives significantly differing results with the TPS-
and the Pad\'e-evaluated analytic parts.
On these grounds,
the obtained optimal $\sigma$ turn out to be
 \label{sig}
\begin{equation}
\sigma  =  0.36 \pm 0.03 \quad (n_f\!=\!4, \ S=1) \ .
\end{equation}

\begin{figure}[htb]
\begin{center}
\begin{minipage}[b]{.49\linewidth}
 \centering\epsfig{file=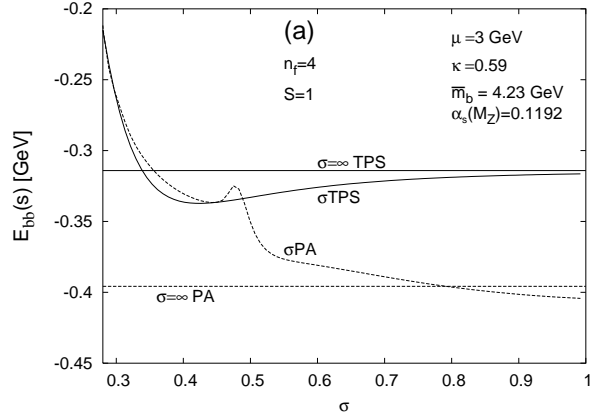,width=\linewidth}
\end{minipage}
\end{center}
 \vspace{0.2cm}
\caption{(a) Soft part of the ground state binding energy of
$b {\bar b}$, evaluated with the (PV) Borel-resummed
expression (\ref{Fssig}), as a function of the
method parameter $\sigma$. Details are given in the text.}
\label{Eqqsig.fig}
\end{figure}

In Fig.~\ref{Eqqsig.fig}(a) we present the   Borel-resummed
soft part of ground state energy for the bottonium ($S=1$),
as a function of the $\sigma$ parameter of method (\ref{Fssig}).

\begin{figure}[htb]
\begin{minipage}[b]{.49\linewidth}
 \centering\epsfig{file=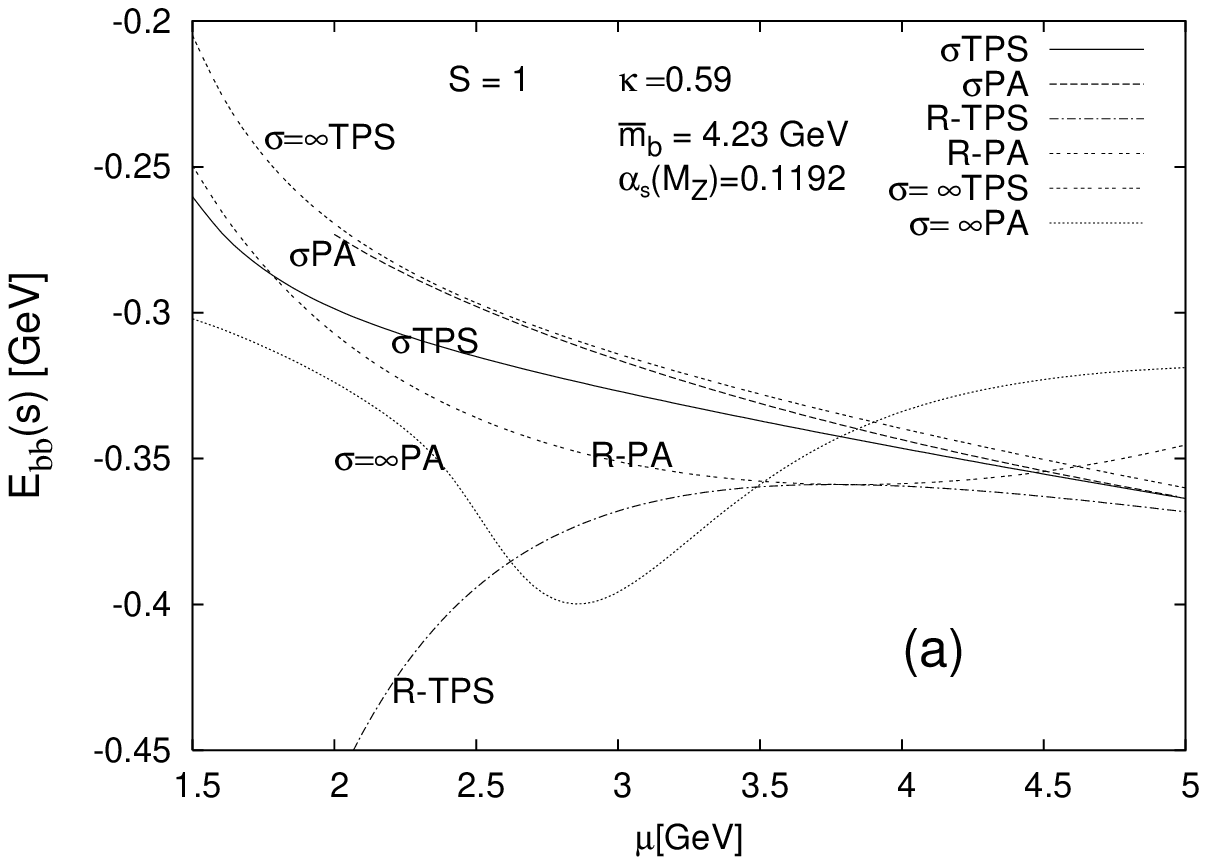,width=\linewidth}
\end{minipage}
\begin{minipage}[b]{.49\linewidth}
 \centering\epsfig{file=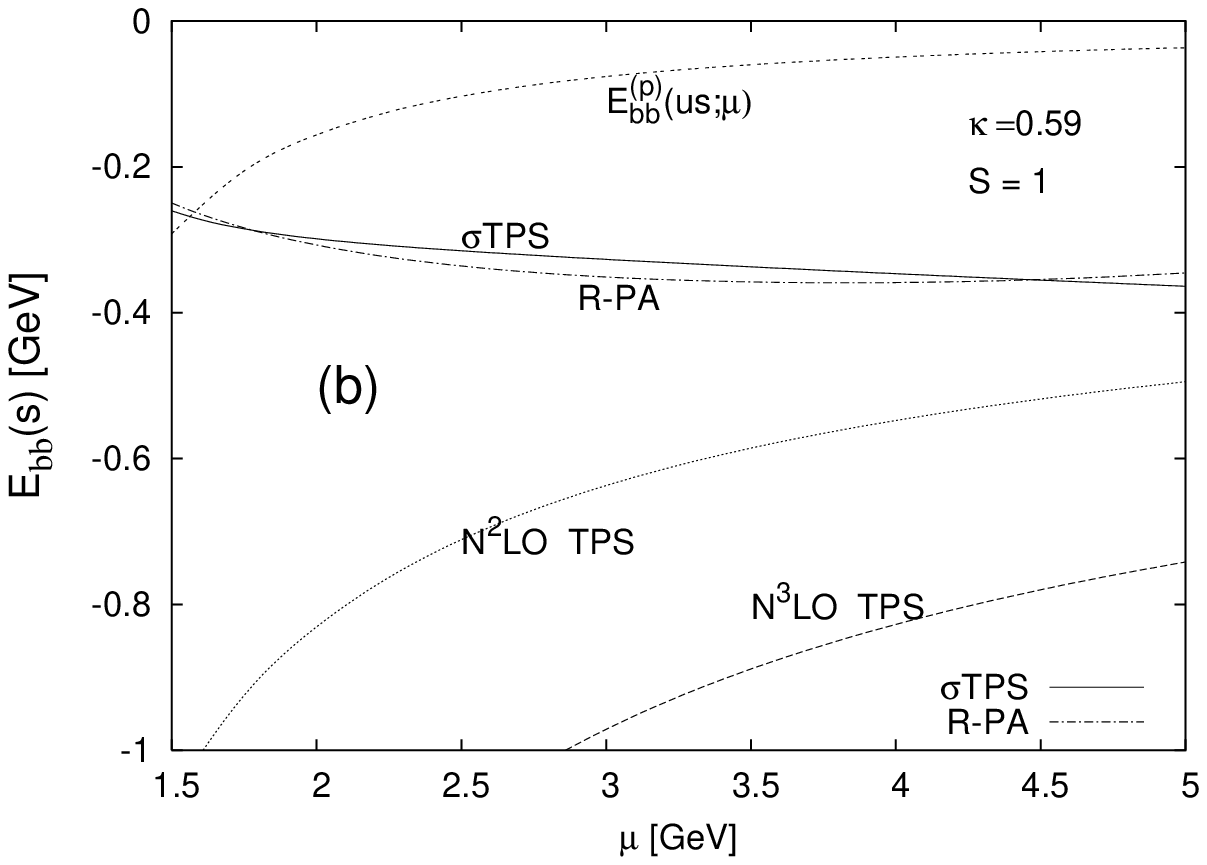,width=\linewidth}
\end{minipage}
\vspace{0.2cm}
\caption{(a) Soft part $E_{b \bar b}(s)$
of the ground state binding energy of
$b {\bar b}$, evaluated with four different methods
involving (PV) Borel resummation, as functions of the
renormalization scale $\mu$. Details are given in the text.
In Fig.~(b) the simple TPS results for $E_{b \bar b}(s)$
are included [Eq.~(\ref{FsTPS})],
as well as the ``perturbative'' ultrasoft part
$E_{b \bar b}^{\rm (p)}(us; \mu)$ [Eq.~(\ref{Fusp})].}
\label{Ebbmu.fig}
\end{figure}

Finally, the results for the soft binding energy $E_{b \bar b}(s)$
of the ground state of
bottonium using  the
$R$-method \cite{Caprini:1998wg,Cvetic:2001sn}, where
we resum the function $R_{F(s)}(b;\mu)$ (\ref{RFs})
and then employ the (PV) Borel resummation as written in
Eq.~(\ref{BRSint}) (with $R_{F(s)}$ instead of $R_S$ there)., as functions of the renormalization scale $\mu$,
are presented in Fig.~\ref{Ebbmu.fig}(a).
We observe from the Figure that the bilocal  ``$\sigma$-regularized'' method (\ref{Fssig}) ($\sigma=0.36$)
gives the TPS and PA results closer to each other.
The methods $\sigma$-TPS, $\sigma$-PA, and $R$-PA give
similar results in the entire presented $\mu$-interval.
$R$-TPS appears to fail at low $\mu$ ($\approx m_b \alpha_s
\approx 1$-$2$ GeV).
In Fig.~\ref{Ebbmu.fig}(b) we include, for comparison,
the simple TPS evaluation of $E_{b \bar b}(s)$, according to
formula [cf.~Eq.~(\ref{Fs})]
\begin{equation}
F(s)^{\rm (TPS)}  \equiv -\frac{9}{4 \pi}
\frac{1}{{\overline m}_b \alpha_s(\widetilde \mu)} E_{q \bar q}(s)
= a(\mu) \left[ 1 + a(\mu) f_1 + a^2(\mu) f_2 + a^3(\mu) f_3 \right]
\ ,
\label{FsTPS}
\end{equation}
where for ${\rm N}^2{\rm LO}$ TPS case we take $f_3=0$.
In Fig.~\ref{Ebbmu.fig}(b) the same input parameters are used
as in Fig.~\ref{Ebbmu.fig}(a).
We see that the perturbation series shows strongly divergent behavior
already at ${\rm N}^3{\rm LO}$.
In this Figure, we also included the ``perturbative''
ultrasoft part $E_{b \bar b}^{\rm (p)}(us;\mu)$ calculated
according to [see Eqs.~(\ref{k30us}) and (\ref{Eqqus})].
\begin{equation}
F^{\rm (p)}(us) \equiv -\frac{9}{4 \pi}
\frac{1}{{\overline m}_b \alpha_s(\widetilde \mu)}
E_{q \bar q}^{\rm (p)}(us;\mu) = k_{3,0} a^4(\mu) \ .
\label{Fusp}
\end{equation}

\subsection{Extraction of bottom mass}
\label{mb}

The estimate of the perturbative part
is given in Eq.~(\ref{Eqqus2}), where it was
essential to take for the renormalization scale
a $us$ scale $\mu \sim \mu_{us} \sim m_q \alpha_s^2$.

For the bottonium case, this scale is below $1$ GeV,
the energy at which we cannot determine perturbatively
$\alpha_s(\mu)$. This indicates that in the bottonium
the $us$ part of the binding energy has an appreciable
nonperturbative part. The lowest energy at which we can
still determine perturbatively $\alpha_s$ is
$\mu \approx 1.5$-$2.0$ GeV,
giving $\alpha_s(\mu) \approx 0.30 - 0.35$.
Although this is a soft scale for $b \bar b$, we will
use this also as an ultrasoft scale. Then by Eq.~(\ref{Eqqus2})
\begin{equation}
E_{b \bar b}(us)^{\rm (p)} \approx
- \frac{4}{9} {\overline m}_q \pi^2 k_{3,0}(us) a^5(\mu_{us})
\approx (- 150 \pm 100) \ {\rm MeV} \ .
\label{Ebbusp}
\end{equation}
The nonperturbative contribution coming from the
gluonic condensate is given by \cite{Voloshin:hc}
\begin{equation}
E_{b \bar b}(us)^{\rm (np)} \approx
{\overline m}_b
\pi^2 \frac{624}{425}
\left( \frac{4}{3} {\overline m}_b \alpha_s(\mu_{us}) \right)^{-4}
\langle a(\mu_{us}) G_{\mu \nu} G^{\mu \nu} \rangle
\approx  (50 \pm 35) \ {\rm MeV} \ ,
\label{Ebbusnp}
\end{equation}
where we used ${\overline m}_b = 4.2$ GeV, and the value
of the gluon condensate $\langle (\alpha_s/\pi) G^2 \rangle =
0.009 \pm 0.007 \ {\rm GeV}^4$  \cite{Ioffe:2002be}.
Eqs.~(\ref{Ebbusp}) and (\ref{Ebbusnp}) give
\begin{equation}
E_{b \bar b}(us)^{\rm (p+np)} \approx (-100 \pm 106) \ {\rm MeV} \ ,
\label{Ebbus}
\end{equation}
The finite charm mass contributions
has been calculated in Ref.~\cite{Brambilla:2001qk}
(Refs.~\cite{Gray:1990yh,Eiras:2000rh,Hoang:2000fm}).
The contribution to the mass
$M_{\Upsilon}(1S) = (2 m_b + E_{b \bar b})$ is
\begin{equation}
\delta M_{\Upsilon}(1S, m_c \not=0) \approx 25 \pm 10 \ {\rm MeV} \ .
\label{mcnot0}
\end{equation}
The estimates (\ref{Ebbus}), and (\ref{mcnot0})
then give a rough estimate of the $us$ and $m_c\!\not=\!0$ contributions
to the bottonium mass
$\delta M_{\Upsilon}(1S; us+m_c) \approx  (- 75 \pm 106)$ MeV.
The mass of the $\Upsilon(1S)$ vector bottonium ground
state is well measured $M_{\Upsilon}(1S) = 9460$ MeV
with virtually no uncertainty \cite{Hagiwara:fs}.
Therefore, the pure perturbative ``soft'' mass is
\begin{equation}
M_{\Upsilon}(1S; s) = 2 m_b + E_{b \bar b}(s) =  9535 \mp 106 \ {\rm MeV}
\ ,
\label{MUs}
\end{equation}
where the uncertainty  is
dominated by the uncertainty of the $us$ regime contribution.
Our numerical results for $E_{b \bar b}(s)$
and for $m_b$   allow us,
by varying the input value of ${\overline m}_b$, to adjust the
sum $2 m_b\!+\!E_{b \bar b}(s)$ to the value given in
Eq.~(\ref{MUs}). For the soft binding energy
we apply the ``$\sigma$-regularized bilocal methods
$\sigma$-TPS and $\sigma$-PA, and $R$-TPS and $R$-PA,
with the  input parameters:
$\alpha_s(M_Z) = 0.1192$; ${\widetilde \mu} = 1.825$ GeV
($\approx \mu_s$), thus $\alpha_s(\widetilde \mu, n_f\!=\!4) =
0.3263$ [$\alpha_s(\mu_s) = 0.326$]; $N_m = 0.555$; $\kappa = 0.59$;
$\sigma = 0.36$; $a_3/4^3 = 86$; $c_4({\overline {\rm MS}}) = 40$.
For $2 m_b$ we apply the bilocal-TPS and $R$-TPS
method, with renormalization scale $\mu_m/{\overline m}_b = 1$,
both methods giving us very similar results
[cf.~Fig.~\ref{mqmu.fig} (a)].
The bilocal-TPS method is applied for $2 m_b$ when $\sigma$-TPS and
$\sigma$-PA are applied for $E_{b \bar b}(s)$; the $R$-TPS
is applied for $2 m_b$ when $R$-TPS and $R$-PA are applied for
$E_{b \bar b}(s)$.
The extracted values
of ${\overline m}_b \equiv
{\overline m}_b(\mu\!=\!{\overline m}_b)$ are then
\begin{eqnarray}
\label{mbb}
{\overline m}_b & = & 4.225 \pm 0.054 \ {\rm GeV} \qquad
(\sigma\!-\!\textrm{TPS}) \ ,
\\
& = & 4.220 \pm 0.056  \ {\rm GeV} \qquad
(\sigma\!-\!\textrm{PA}) \ ,
\\
& = & 4.243 \pm 0.080  \ {\rm GeV} \qquad
(R\!-\!\textrm{TPS}) \ ,
\\
& = & 4.235 \pm 0.068  \ {\rm GeV} \qquad
(R\!-\!\textrm{PA}) \ .
\end{eqnarray}
Finally, the  average
of the central values  gives us
\begin{equation}
{\overline m}_b = 4.231 \pm 0.068 \ {\rm GeV}  \ .
\label{mbbav}
\end{equation}

\section{Comparisons and conclusions}
\label{summ}

Our result for the mass ${\overline m}_{b}=4.231 \pm 0.068 \ \rm GeV$,
is compared in Table \ref{table7}  with   recently values of ${\overline m}_b$.
The only input parameter common to all these
methods is $\alpha_s$.
Our central value was $\alpha_s(m_{\tau}) = 0.3254$
[$\Rightarrow \alpha_s(M_Z) = 0.1192$] since such \cite{Cvetic:2001ws},
or similar \cite{Geshkenbein:2001mn,Cvetic:2001sn},
values follow from the
(nonstrange) semihadronic $\tau$ decay data which are
very precise \cite{Barate:1998uf}. On the other hand, the
world average as of September 2002 is
$\alpha_s(M_Z) = 0.1183 \pm 0.0027$ \cite{Bethke:2002rv}.
Most of the authors during the last four years used
central value $\alpha_s(M_Z) \approx 0.118$.
Therefore, for comparisons, we convert our
results (\ref{mbb}) to this central value of $\alpha_s$ --
more specifically, from $\alpha_s(M_Z) = 0.1192 \pm 0.0015$
to $0.1180 \pm 0.0015$. This can be easily done by
inspecting in Table \ref{table2} the column
under $\alpha_s$,
an increase in the central values of ${\overline m}_b$ by $11$, $12$, $8$ and $10$ MeV,
respectively. This gives the average $10$ MeV
higher than in Eq.~(\ref{mbbav})
\begin{equation}
{\overline m}_b = 4.241 \pm 0.068 \ {\rm GeV} \quad
 \textrm{average \  when:} \  \alpha_s(M_Z) = 0.1180 \pm 0.0015   \ .
\label{mbbav2}
\end{equation}

There are two important numerical effects in our result.
The first is the separate evaluation of the ``perturbative''
ultrasoft energy part at the corresponding low renormalization energy
($ 1.5 - 2$ GeV), Eqs.~(\ref{Eqqus2}) and (\ref{Ebbusp}).
If we had not separated the (``perturbative'') ultrasoft from
the soft part of the binding energy,
the use of the common renormalization
energy scale $\mu$ ($\approx 3$ GeV) in the resummation then
would have given us the central value of $E_{b \bar b}(us)$
by about $+100$ MeV higher. Then the
extracted value of ${\overline m}_b$
would have gone down by about $46$ MeV, giving the
value ${\overline m}_b \approx 4.195 \pm 0.068$,
with the central value close to that of L03 in Table \ref{table7}.
On the other hand, that latter value is quite clearly
lower than the value PS02 in Table \ref{table7}, by about $150$ MeV,
principally because of the $b=1/2$ renormalon effect which
were taken into account here and in Ref.~\cite{Lee:2003hh}.
Thus, the renormalon effect brings down the extracted central
value of ${\overline m}_b$ by about 150 MeV, but the separate
evaluation  of the ultrasoft contribution brings
it up by about $50$ MeV.

\section*{Acknowledgments}
I would like  to thank G. Cvetic for helpful and discussions. This work was supported  by  Project USM-110321 of the UTFSM.
\appendix

\section{Coefficients for the expansion of the soft binding energy}
\label{app:sbe}

We write down here the explicit coefficients $f_j$ of
the expansion (\ref{Fs}) for the soft part of the
ground state binding energy. The logarithms appearing
in these expressions involve three scales
[$\mu, {\widetilde \mu}, {\overline m}_q$ and
${\overline \mu}(\widetilde \mu) =
(4/3) {\overline m}_q \alpha_s(\widetilde \mu)$]
\begin{eqnarray}
L_1 & = & \ln \left(
\frac{{\overline m}_q}{{\overline \mu}(\widetilde \mu)} \right) \ ,
\qquad
L_2 = \ln \left( \frac{ {\overline m}_q}{\widetilde \mu} \right) \ ,
\qquad
L_{\mu} = \ln \left( \frac{ {\overline m}_q}{\mu} \right) \ .
\label{logs}
\end{eqnarray}
The coefficients $f_j$ are
\begin{eqnarray}
f_1 & = & \frac{1}{2} (35 + 22 L_1 - 11 L_{\mu} - 11 L_2 )
+ \frac{1}{9} (-11 - 6 L_1 + 3 L_{\mu} + 3 L_2) n_f \ .
\label{f1}
\end{eqnarray}
 \label{f2}
\begin{eqnarray}
f_2 & = & f_{2}^{(0)} + f_2^{(1)} n_f + f_2^{(2)} n_f^2 \ ,
\label{f2exp}
\\
f_{2}^{(0)}  & = & {\big [}
381.67 + 90.75 L_1^2  + 30.25 L_{\mu}^2
+ L_1 (246.42 - 121 L_{\mu} - 60.5 L_2) - 48.5 L_2
\nonumber\\
&&
+ L_{\mu} (-205.25 + 60.5 L_2) - 11.697 S(S+1) {\big ]}\nonumber  \ ,
\label{f20}
\\
f_{2}^{(1)}  & = & {\big [}
-42.7469 - 11 L_1^2  - 3.6667 L_{\mu}^2
+ L_{\mu} (26.6944 - 7.3333 L_2) + 6.8056 L_2
\nonumber\\
&&
+ L_1 (-33.0556 + 14.6667 L_{\mu} + 7.3333 L_2) {\big ]}\nonumber  \ ,
\label{f21}
\\
f_{2}^{(2)}  & = &  {\big [}
1.16286 + (3/9) L_1^2  + (1/9) L_{\mu}^2
+ L_1 (1 - (4/9) L_{\mu} - (2/9) L_2)
\nonumber\\
&&
+ L_{\mu} (-0.81482 + (2/9) L_2) - 0.18518 L_2 {\big ]} \nonumber \ .
\label{f22}
\end{eqnarray}
\label{f3}
\begin{eqnarray}
f_3 & = & f_{3}^{(0)} + f_3^{(1)} n_f + f_3^{(2)} n_f^2
+ f_3^{(3)} n_f^3 \ ,
\label{f3exp}
\\
f_{3}^{(0)}  & = & {\big [}
6726.11 + 665.5 L_1^3
- 166.375 L_{\mu} \left( 40.802 + (-10.599 + L_{\mu}) L_{\mu} \right)
\nonumber\\
&&
+ L_1^2  (2381.5 - 1497.38 L_{\mu} - 499.125 L_2) - 871.429 L_2
\nonumber\\
&& - 499.125 (-1.884 + L_{\mu}) L_{\mu} L_2
 - 201.438 L_2^2 + L_1  ( 7457.17 \nonumber\\
&& - 497.292 L_2 +
L_{\mu} (-4346.38 + 998.25 L_{\mu} + 998.25 L_2)  )
\nonumber\\
&&
- 257.341 (0.2112 +  L_1 - 0.75 L_{\mu} - 0.25 L_2) S (S+1)
\nonumber\\
&&
- 61.4109 \left( -6.1394 + S (S+1) \right)
\ln \left( \alpha_s(\mu_s) \right) + 440.172 \ln ( \kappa )
+ 2 a_3/4^3
{\big ]}\nonumber  \ ,
\label{f30}
\\
f_{3}^{(1)}  & = & {\big [}
-1274.33 - 1277.92 L_1 - 471.125 L_1^2  - 121 L_1^3  + 1182.32 L_{\mu}
\nonumber\\
&&
+ 843.667 L_1 L_{\mu} + 272.25 L_1^2  L_{\mu}
 - 335.813 L_{\mu}^2  - 181.5 L_1 L_{\mu}^2 \nonumber\\
&&
+ 30.25 L_{\mu}^3  + 124.501 L_2 + 108.361 L_1 L_2 + 90.75 L_1^2  L_2
\nonumber\\
&&
-186.708 L_{\mu} L_2
- 181.5 L_1 L_{\mu} L_2 + 90.75 L_{\mu}^2  L_2 + 36.729 L_2^2
\nonumber\\
&&
+ (4.0686 + 15.5964 L_1  - 11.6973 L_{\mu}  - 3.8991 L_2) S (S+1)
{\big ]} \nonumber \ ,
\label{f31}
\\
f_{3}^{(2)}  & = & {\big [}
70.8992 + 70.2453 L_1 + 28.9722 L_1^2  + 7.3333 L_1^3  - 65.9925 L_{\mu}
\nonumber\\
&&
-51.6667 L_1 L_{\mu}
- 16.5 L_1^2  L_{\mu} + 20.5972 L_{\mu}^2 + 11  L_1 L_{\mu}^2
-1.8333 L_{\mu}^3 \nonumber\\
&& - 5.1939 L_2 - 6.5741 L_1 L_2
- 5.5 L_1^2  L_2 + 10.9167 L_{\mu} L_2 \nonumber\\
&& + 11  L_1 L_{\mu} L_2
- 5.5 L_{\mu}^2  L_2 - 2.0972  L_2^2
{\big ]} \nonumber \ ,
\label{f32}
\\
f_{3}^{(3)}  & = & {\big [}
-1.21475 - 1.21714 L_1 - (5/9) L_1^2  - 0.14815 L_1^3
+  1.16286 L_{\mu} \nonumber\\
&& + L_1 L_{\mu}
+ (1/3) L_1^2  L_{\mu}
- 0.40741 L_{\mu}^2
- (2/9) L_1 L_{\mu}^2  + 0.03704 L_{\mu}^3 \nonumber\\
&& + 0.05429 L_2
+ (1/9) L_1 L_2 + (1/9) L_1^2  L_2
\nonumber\\
&&
- 0.18518 L_{\mu} L_2 - (2/9) L_1 L_{\mu} L_2
+ (1/9) L_{\mu}^2  L_2 + 0.03704 L_2^2
 {\big ]} \nonumber \ .
\label{f33}
\end{eqnarray}

\newpage


\newpage

\begin{table}
\caption{\label{table2} The separate uncertainties
$\delta {\overline m}_b$ (in MeV) for the extracted
value of ${\overline m}_b$ from various sources:
1.) $us$ [$\delta E_{b \bar b}(us)^{\rm (p+np)} = -100 \pm 106$ MeV];
2.) $\mu = 3 \pm 1$ GeV;
3.) $\mu_m = {\overline m}_b ( 1 \pm 0.5)$;
4.) $\alpha_s(m_{\tau}) = 0.3254 \pm 0.0125$
[$\alpha_s(M_Z) = 0.1192 \pm 0.0015$];
5.) $N_m = 0.555 \pm 0.020$ [$\kappa = 0.59 \pm 0.19$];
6.) $a_3/4^3 = 86.\pm 23.$;
7.) $c_4 = 40. \pm 60.$;
8. $\sigma = 0.36 \pm 0.03$;
9. $m_c \not= 0$ ($\delta M_{\Upsilon}(m_c\!\not=\!0)=\pm 10$ MeV).}
\begin{center}
\begin{tabular}{crrrrrrrrr}
 & $us$ & $\mu$ & $\mu_m$ & $\alpha_s$ & $N_m$ & $a_3$ & $c_4$ &
$\sigma$ & $m_c$ \\
\hline
$\sigma$-TPS & $- 49$ & $+ 9$ & $- 4$ & $-13$ &$ -3$  &$ + 2$ &
$ - 8$ & $+ 4$ & $- 5$
\\
 & $+ 49$ & $- 13$ & $+ 2$ &$ + 14$ &$ + 2$ &$ - 2$ & $+ 8$ &
$ - 9$ &$ + 5$
\\
\hline
$\sigma$-PA &  $- 49 $ & $ + 13 $ &  $- 4 $ & $ - 15 $ &
$ -3 $ & $ + 1 $ & $ - 5 $ & $ + 5 $ & $ -  5$
\\
 & $ + 49 $ & $ -20 $ & $ + 2 $ & $ + 15 $ & $ + 2 $ &
 $- 1 $ & $ + 2 $ & $ - 9 $ & $ + 5 $
\\
\hline
$R$-TPS & $ - 50 $ & $ -4 $ & $ + 4 $ & $ - 8 $ &
 $ - 9 $ & $ - 3 $& $0$ & $0$  &  $-5$
\\
 &  $+50 $ & $ + 45 $ & $ - 40 $ & $ + 10 $ &
 $ + 11 $ & $ + 3 $ &$0$ & $0$  &  $+5$
\\
\hline
$R$-PA &  $- 49 $& $ + 3 $ & $ + 4 $ & $ - 11 $ &
 $ -4 $  & $- 2 $  &$0$  &$0$  &  $-5$
\\
 &  $+ 49 $ & $-20 $ &  $- 40 $ &  $+ 12 $ &
 $+ 4 $ &  $+ 2 $ &$0$ &$0$ &  $+5$
\\
\end{tabular}
\end{center}
\end{table}

\begin{table}
\caption{\label{table7} Recently obtained values of
(${\overline {\rm MS}}$) ${\overline m}_b$ mass obtained from
$\Upsilon$ sum rules or from spectrum of the
$\Upsilon$(1S) resonance. Wherever needed
(\cite{Penin:2002zv,Lee:2003hh}),
the central mass values were adjusted to the
common input central value $\alpha_s(M_Z) = 0.118$.}
\begin{center}
\begin{tabular}{l c l l}
reference & method & order & ${\overline m}_b({\overline m}_b)$ (GeV)
\\
\hline
PP98 \cite{Penin:1998zh} & $\Upsilon$ sum rules & NNLO &
$4.21 \pm 0.11$
\\
MY98 \cite{Melnikov:1998ug} & $\Upsilon$ sum rules & NNLO &
$4.20 \pm 0.10$
\\
BS99 \cite{Beneke:1999fe}  & $\Upsilon$ sum rules & NNLO &
$4.25 \pm 0.08$
\\
H00 \cite{Hoang:2000fm} & $\Upsilon$ sum rules & NNLO &
$4.17 \pm 0.05$
\\
KS01 \cite{Kuhn:2001dm}  & $\Upsilon$ sum rules & NNLO &
$4.209 \pm 0.050$
\\
CH02 \cite{Corcella:2002uu} & $\Upsilon$ sum rules & NNLO &
$4.20 \pm 0.09$
\\
E02 \cite{Eidemuller:2002wk} &  $\Upsilon$ sum rules & NNLO &
$4.24 \pm 0.10$
\\
P01 \cite{Pineda:2001zq} & spectrum, $\Upsilon$(1S) & NNLO &
$4.210 \pm 0.090 \pm 0.025$
\\
BSV01 \cite{Brambilla:2001qk}  & spectrum, $\Upsilon$(1S) & NNLO &
$4.190 \pm 0.020 \pm 0.025$
\\
PS02 \cite{Penin:2002zv}  & spectrum, $\Upsilon$(1S) &${\rm N}^3{\rm LO}$ &
$4.349 \pm 0.070$
\\
L03 \cite{Lee:2003hh} & spectrum, $\Upsilon$(1S) &${\rm N}^3{\rm LO}$ &
$4.19 \pm 0.04$
\\
this work, Eq.~(\ref{mbbav2})&spectrum, $\Upsilon$(1S)&${\rm N}^3{\rm LO}$&
$4.241 \pm 0.070$
\\
\end{tabular}
\end{center}
\end{table}

\end{document}